\documentclass[showpacs,aps,superscriptaddress,twocolumn,floatfix]{revtex4}
\usepackage{graphicx,float,amsmath,amssymb}

\begin{document}

\title{Plasmon modes of silver nanowire on a silica substrate}
\author{C.-L. Zou}
\affiliation{Key Lab of Quantum Information, University of Science
and Technology of China, Hefei 230026, Anhui, P. R. China}
\author{F.-W. Sun}
\affiliation{Key Lab of Quantum Information, University of Science
and Technology of China, Hefei 230026, Anhui, P. R. China}
\email{fwsun@ustc.edu.cn}
\author{Y.-F. Xiao}
\affiliation{State Key Lab for Mesoscopic Physics, School of
Physics, Peking University, Beijing 100871, P. R. China}
\author{C.-H.Dong}
\affiliation{Key Lab of Quantum Information, University of Science
and Technology of China, Hefei 230026, Anhui, P. R. China}
\author{X.-D. Chen}
\affiliation{Key Lab of Quantum Information, University of Science
and Technology of China, Hefei 230026, Anhui, P. R. China}
\author{J.-M. Cui}
\affiliation{Key Lab of Quantum Information, University of Science
and Technology of China, Hefei 230026, Anhui, P. R. China}
\author{Q. Gong}
\affiliation{State Key Lab for Mesoscopic Physics, School of
Physics, Peking University, Beijing 100871, P. R. China}
\author{Z.-F. Han}
\affiliation{Key Lab of Quantum Information, University of Science
and Technology of China, Hefei 230026, Anhui, P. R. China}
\email{zfhan@ustc.edu.cn}
\author{G.-C. Guo}
\affiliation{Key Lab of Quantum Information, University of Science
and Technology of China, Hefei 230026, Anhui, P. R. China}

\begin{abstract}
Plasmon mode in a silver nanowire is theoretically studied when the
nanowire is placed on or near a silica substrate. It is found that
the substrate has much influence on the plasmon mode. For the
nanowire on the substrate, the plasmon (hybrid) mode possesses not
only a long propagation length but also an ultrasmall mode area.
From the experimental point of view, this cavity-free structure
holds a great potential to study a strong coherent interaction
between the plasmon mode and single quantum system (for example,
quantum dots) embedded in the substrate.
\end{abstract}

\maketitle

The silver nanowire is a potential solution to key components in
future ultra-compact electronic and photonic circuit
\cite{ozbay,wireexp2,wireexp3} since it can confine the light in
nanoscale beyond the diffraction limit. The strong confinement of
light in a small area produces strong electric field intensity and
dramatically changes the density of state nearby the metal surface.
Therefore, the silver nanowire has been suggested to realize strong
light-matter interaction, where the nanowire plays as the role of a
resonator in cavity quantum electrodynamics (QED). When an optical
emitter, e.g., a quantum dot (QD), is placed around the silver
nanowire, its spontaneous emission can be significantly
modified\cite{chang1}, known as the Purcell effect. As a result,
single optical emitter coupling with plasmon
modes of nanowire holds a great potential for broad-band cavity QED \cite%
{chang1}, single photon source \cite{akimov}, and sub-wavelength
single photon transistor \cite{chang2,chen} for quantum information
science.

Theoretically, the plasmon mode in silver nanowire was studied as an
ideal axis-symmetric mode \cite{chang1}, or still treated without
the
external influence even it was placed on a substrate \cite%
{zma,yma,tshegai,rxyan,dong}. However, in realistic applications,
the silver nanowire is always necessary to be located on or near a
substrate. For example, the nanowire on a substrate makes it
convenient for manipulating assisted by the
scanning near field optical microscope probe \cite%
{zma,yma,tshegai,rxyan,pyayt,dong,sanders,hwei,baik}. To the best of
our knowledge, the influence on the plasmon mode from the substrate
is still kept un-studied. Actually, the plasmon mode is very
sensitive to the surrounding environment. In this Letter, we
theoretically study the plasmon mode propagating along the silver
nanowire near a substrate with a gap $g\geq 0$. In this
silver-air-substrate structure, most of the mode energy is confined
around the interface between the substrate and silver nanowire. It
is very different from the nanowire simply embedded in air. The
benefit of the nanowire on a substrate is that the large portion of
energy in the substrate reduces the mode area by a factor of 5,
without significant degradation of the propagation loss.

\begin{figure}[ptb]
\centerline{
\includegraphics[width=6.3cm]{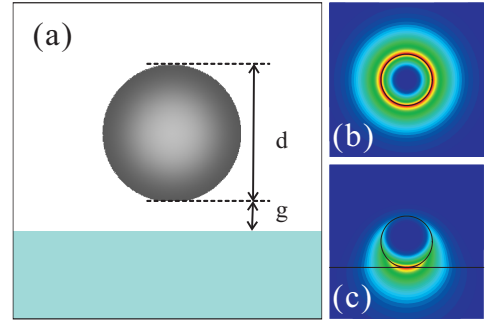}}
\caption{(color online) (a) Schematic diagram of a silver nanowire
near the silica substrate with a gap $g$. The diameter of nanowire
is $d$. (b) and (c), the energy density distributions on the cross
section of nanowire ($d=100nm$) with $g=\infty,0$, respectively.}
\end{figure}

An infinite long silver nanowire with a diameter of $d$ is near an
infinite dielectric substrate with a gap of $g$. The cross section
is shown Fig. 1(a). The surface plasmon mode propagates harmonically
along the nanowire,
and the electric field varies as $\mathrm{exp}(\mathrm{i}\beta z-\mathrm{i}%
\omega t)$, where the propagation constant $\beta =2\mathrm{\pi }%
n_{eff}/\lambda $ is a complex number because of the energy
attenuation in the propagation. In this case, the Maxwell equation
is reduced to two-dimensional cross section, which can be expressed
as
\begin{equation}
\lbrack \bigtriangledown ^{2}+(n^{2}-n_{eff}^{2})(2\pi /\lambda )^{2}]\psi =0%
\text{,}
\end{equation}%
where $n$ and $n_{eff}$ denote the refractive index of material and
the effective index for the mode, respectively. The propagation
length of the plasmon modes can be defined as
\begin{equation}
L=1/2\text{\textrm{Im}}\{\beta \}\text{.}
\end{equation}

Besides the propagation loss, another important parameter of the
plasmon mode is effective mode area $A$, which can be defined by the
ratio of a mode's total energy density per unit length and its peak
energy density,
\begin{equation}
A=\int_{all}W(r)\mathrm{d}s/max\{W(r)\}\text{,}
\end{equation}%
where $W(r)$ represents the effective energy density with
\begin{equation}
W(r)=\frac{1}{2}\mathrm{Re}\{\frac{\mathrm{d}[\omega \varepsilon (r)]}{%
\mathrm{d}\omega }\}\left\vert E(r)\right\vert ^{2}+\frac{1}{2}\mu
_{0}\left\vert H(r)\right\vert ^{2}\text{.}
\end{equation}%
Here, $\left\vert E(r)\right\vert ^{2}$ and $\left\vert
H(r)\right\vert ^{2}$
are the intensity of electric and magnetic fields, respectively. $%
\varepsilon (r)$ and $\mu _{0}$ are the electric and vacuum magnetic
permittivities.

To describe the impact from the substrate, we also define the
confinement factor $\eta $ as the portion of energy in substrate,
\begin{equation}
\eta =\int_{sub}W(r)\mathrm{d}s/\int_{all}W(r)\mathrm{d}s\text{.}
\end{equation}%
The confinement factor $\eta $ can also quantify the interaction of
SPP and substrate,

A finite element method is used to investigate the plasmon modes
with a fixed wavelength at $637$ nm. The energy distribution
profiles of the plasmon modes in the nanowire are shown in Fig. 1(b)
and (c), with $g=\infty ,0$, respectively. Here, $n_{air}=1$ and
$n_{sub}=1.45$. It is not difficult
to find a rotation symmetric field distribution when the nanowire is in air (%
$g=\infty $), while a strongly asymmetric field distribution is
presented in the presence of the silica substrate ($g=0$). Moreover,
the maximum of the electric field is distributed in the
substrate-nanowire gap. In other words, the substrate can
significantly pull the mode field.

\begin{figure}[ptb]
\centerline{
\includegraphics[width=6.3cm]{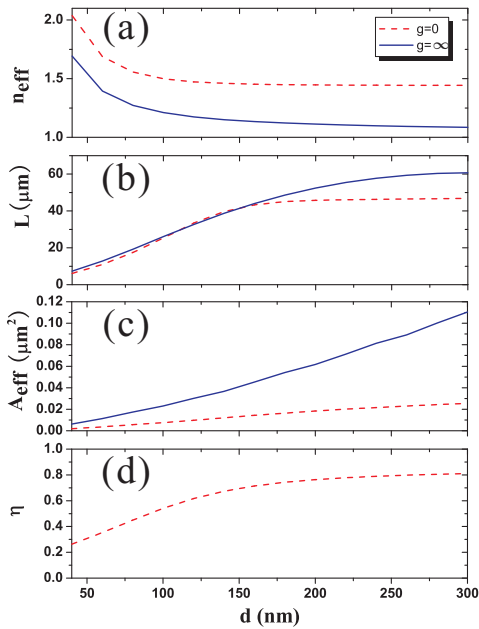}}
\caption{(color online) Effective index (a), propagation loss (b),
mode areas (c), and energy confinement in the substrate (d) of
surface plasmon modes for the nanowire on the substrate (red dashed,
$g=0$) and in air (blue solid, $g=\infty$) versus the diameter of
silver nanowire.}
\end{figure}

We first investigate the impact of the substrate on the plasmon
modes for various nanowire sizes. Fig.2 shows the effective index
($n_{eff}$), propagation length ($L$), effective mode area
($A_{eff}$) and portion of energy ($\eta $) versus $d$. As a
comparison, the ideal case without substrate ($g=\infty $) is also
depicted. It is found that $n_{eff}$ decreases with the increase of
$d$, while $L$ and $A_{eff}$ shows the opposite behaviors. This
tradeoff between confinement and absorption loss is well known in
the surface plasmon mode.

For the nanowire on silica substrate, comparing with the nanowire in air, $%
n_{eff}$ is larger and $A_{eff}$ is smaller with all $d$. These
obvious differences come from the presence of large refractive index
material near the metal nanowire. In addition, the mode area of the
nanowire on silica substrate is only about 1/5 of that in air. To
characterize the field pulling by the silica substrate, we present
the confinement factor for $g=0$ in Fig.2(d). The $\eta $ increases
with $d$ to nearly $80\%$ when $d=200nm$.

In Fig.(2), $A_{eff}$ simply increases with the increase of $d$.
However, the other three parameters approach saturation values. In
order to get a physic interpretation, we can approximately treat
this hybrid system as the metal plane-substrate for large $d$. The
analytical solution to Eq. (1) in the plane condition gives
$n^{p}{}_{\text{eff}}=\sqrt{\epsilon _{m}\epsilon _{s}/\left(
\epsilon _{m}+\epsilon _{s}\right) }$, where $\epsilon _{m}$ and
$\epsilon _{s}$ is the electric permittivity of metal and substrate.
So we have $n^{p}{}_{\text{eff}}=1.537$, $L^{p}=53\mu m$, and $\eta
^{p}=90\%$ in the plane limit.

Now, we turn to discuss the experimental condition, where the
nanowire is near the substrate with a finite gap $g>0$. This is a
realistic case for many experiments because the substrate is not
perfect smooth, or in specific experimental config, the nanowire is
suspended in air\cite{falk}. By fixing the diameter $d=100$ nm, we
study the propagation loss and effective index (Fig.3) for different
gaps between nanowire and substrate.

With the increase of gap $g$, $n_{eff}$ monotonously decreases and
gradually approaches the in-air case. This is because the influence
of substrate becomes smaller when the nanowire departs from the
substrate. When $g=200$ nm, the substrate has a very minor effect on
the plasmon mode of the silver.
However, the dynamics of $L$ is much more complicated than that of $%
n_{eff}$. When the silver nanowire moves away from the silica
substrate, $L$ increases first, then quickly drops to $2$ $\mu m$.
After
that, it slowly increases, and finally approaches to $25$ $\mu $m \cite{note}%
. We can divide the gap into three regimes: the near field,
dissipation, and far field regimes, which correspond to different
dynamics of $L$ and different loss mechanisms.

\begin{figure}[ptb]
\centerline{
\includegraphics[width=7.3cm]{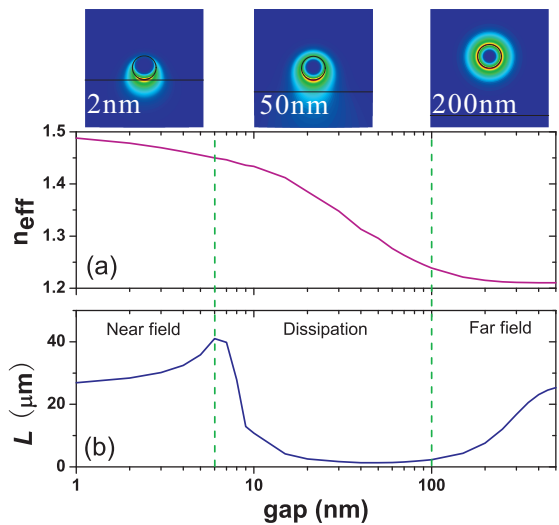}}
\caption{(color online) Effective index (a) and propagation length
(b) of the silver nanowire with d=100 nm versus the gap. Insets:
mode field profiles of the surface plasmon modes with g=2 nm, 50 nm,
200 nm, respectively.}
\end{figure}

(i) Near field regime. In this regime, we have $n_{eff}>n_{sub}$.
Due to the phase mismatching, the dominant loss of the plasmon mode
comes from the metal absorption. By increasing the gap, $n_{eff}$
gradually decreases as the nanowire moves away from the substrate. A
smaller $n_{eff}$ indicates a smaller energy portion in the metal,
so that the absorption loss is reduced with increasing $g$.

(ii) Dissipation Regime. When the gap further increases, $n_{eff}$
may be smaller than $n_{sub}$. Meanwhile, the plasmon mode still has
much energy in the substrate, as can be seen in the inset of Fig.3
($g=50$ nm). From Eq. (1), when $n_{eff}<n_{sub}$, the solutions are
traveling waves in substrate as $\psi \varpropto \exp
(\mathrm{i}\sqrt{n_{sub}^{2}-n_{eff}^{2}}x)$. The energy in
traveling wave will dissipate into infinite space. Therefore, in
this regime, the mode energy quickly dissipates by coupling to the
dissipative traveling waves in the substrate. As a result, $L$ has a
sudden decrease.

(iii) The far field regime. When the nanowire further moves away
from the substrate, for example, $g=200$ nm, the coupling between
the plasmon mode and the dissipative traveling modes in substrate
becomes very weak as it is proportional to their field overlap. Thus, even $%
n_{eff}<n_{sub}$, the plasmon mode keeps almost unchanged by the
substrate, and the mode profile approaches the case of $g=\infty $
(in the absence of the substrate). Therefore, $L$ can be slowly
recovered in this regime.

The above results indicate that the approximation by neglecting the
substrate is not applicable in analyzing experimental results, where
the energy confinement and propagation loss are significantly
influenced by the substrate. When the gap is very small, most of the
energy is confined near nanowire-substrate interface and $n_{eff}$
is large. It should be more efficient to coupling light to the
surface plasmon mode through the substrate, especially for large
$d$. The efficiently excitation of surface plasmon mode through
substrate have been demonstrated recently\cite{tshegai,rxyan}. It is
worth noticing that the propagation loss strongly depends on the
gap. In a recent experiment, near field surface plasmon mode
detection is demonstrated\cite{falk}, where the silver nanowire is
put on the Ge nanowire, with $g=40$nm, $d=100$nm and $\lambda
=600$nm. From Fig. 3(b), it works in the dissipation regime with
$L\approx 2\mu m$. For more efficiently detection, $g$ should be
larger than $200$nm or smaller than $10$nm .

In conclusion, we have numerically studied the mode profile,
propagation loss and mode area of plasmon modes of a silver nanowire
near the silica substrate. To reduce the propagation loss, our
result suggests that the gap between silver nanowire and the
substrate should be as small as possible. When the silver nanowire
is attached to the substrate ($g=0$), the propagation length is
limited by metal absorption and the mode area is reduced by a factor
of $5$. In addition, for nanowire with $d=200$ nm, the
field energy portion of the plasmon mode in the substrate is as large as $%
80\%$, which is suitable for coupling to QDs or molecules embed in
substrate. Accompanying with the scanning near field optical
microscope, the silver nanowire can be precisely moved on the
substrate, and selectively coupled to single QDs in substrate. This
is potential for experimental realization of the single photon
transistors, or coupling two distant QDs to generate entanglement
for quantum information science.

The work was supported by the National fundamental Research Program
of China under Grant No 2006CB921900, the Knowledge Innovation
Project of Chinese Academy of Sciences, and National Natural Science
Foundation of China (Grant No. 60621064, 60537020). F.W. S. is also
supported by the new faculty starting funds from USTC and the
Fundamental Research Funds for the Central Universities.

\newpage


\begin{thebibliography}{99}
\bibitem{note} Similar behavior of SPP have been observed in the case of
sliver slab near a substrate, which could be solved analytically.
\end{thebibliography}
\end{document}